\documentclass{aa}

\usepackage{graphicx}
\usepackage{natbib,twoopt}
\usepackage[breaklinks=true]{hyperref} 
 \bibpunct{(}{)}{;}{a}{}{,}    
 \newcommandtwoopt{\citeads}[3][][]{\href{http://adsabs.harvard.edu/abs/#3}%
                                        {\citealp[#1][#2]{#3}}}
 \newcommandtwoopt{\citepads}[3][][]{\href{http://adsabs.harvard.edu/abs/#3}%
                                        {\citep[#1][#2]{#3}}}
 \newcommandtwoopt{\citetads}[3][][]{\href{http://adsabs.harvard.edu/abs/#3}%
                                        {\citet[#1][#2]{#3}}}
 \newcommandtwoopt{\citeyearads}[3][][]%
   {\href{http://adsabs.harvard.edu/abs/#3}{\citeyear[#1][#2]{#3}}}
\bibpunct{(}{)}{;}{a}{}{,}
\usepackage{latexsym}
\usepackage{subfigure}
\usepackage{amsmath}
\usepackage{amssymb}
\usepackage{amstext}
\usepackage{color}
\usepackage{psfrag}
\usepackage{txfonts}

\def\cm3{cm$^{-3}$}

\def\12{$^{12}$CO}

\usepackage{graphicx}
\usepackage{txfonts}
\begin{document}

\title{A hadronic scenario for HESS~J1818$-$154}

\author{G. Castelletti\inst{1}
\and L. Supan\inst{1}
\and G. Dubner\inst{1}
\and B.~C. Joshi\inst{2} 
\and M.~P. Surnis\inst{2}
}

\offprints{G. Castelletti}
\institute {Instituto de Astronom\'{\i}a y  F\'{\i}sica del Espacio (IAFE), UBA-CONICET,
CC 67, Suc. 28, 1428 Buenos Aires, Argentina\\
             \email{gcastell@iafe.uba.ar}
\and National Centre for Radio Astrophysics (NCRA), Pune, India }

 \date{Received <date>; Accepted <date>}

\abstract
   {}
   {\object{G15.4+0.1} is a faint supernova remnant (SNR) that has recently been associated with the $\gamma$-ray source \object{HESS~J1818$-$154}. We investigate a hadronic
scenario for the production of the $\gamma$-ray emission.}
   {Molecular $^{13}$CO (J=1$-$0) taken from the Galactic Ring Survey (GRS)
and neutral hydrogen (HI) data from the Southern 
Galactic Plane Survey (SGPS) have been used in combination
with new 1420~MHz radio continuum observations carried out with the Giant Metrewave Radio Telescope (GMRT).}
   {From the new observations and analysis of archival data we provided for the first time a
reliable estimate for the distance to the SNR~G15.4+0.1 and discovered molecular clouds located at the same distance.
On the basis of HI absorption features, we estimate the distance to G15.4+0.1 in 4.8$\pm$1.0~kpc.
The $^{13}$CO observations clearly show a molecular cloud about 5$^{\prime}$ in size 
with two bright clumps, labeled A and B, clump A positionally associated with the location
of HESS~J1818$-$154 and clump B in coincidence with the brightest northern
border of  the radio SNR shell. The HI absorption and the $^{13}$CO emission study indicates a possible interaction between 
the molecular material and the remnant. We estimate the masses and densities of the molecular gas as
(1.2$\pm$0.5)$\times$10$^{3}$M$_{\odot}$ and (1.5 $\pm$ 0.4)$\times$10$^{3}$~cm$^{-3}$ for clump A and
(3.0$\pm$0.7$)\times$10$^{3}$M$_{\odot}$ and (1.1$\pm$0.3)$\times$10$^{3}$~cm$^{-3}$ for clump B.
Calculations show that the average density of the molecular clump A is sufficient to produce
the detected $\gamma$-ray flux, thus favoring a hadronic origin for the high-energy emission.}
{}


\maketitle

\titlerunning{The G15.4+0.1/HESS~J1818$-$154 system}
\authorrunning{Castelletti et al.}
%

\section{Introduction}

The source G15.4+0.1 was first identified as a faint supernova remnant (SNR) 
during a survey of the Galactic plane carried out with the Very Large Array (VLA) at 330~MHz 
\citep{bro06}. It is characterized by an almost circular shell with a diameter
of nearly 15$^{\prime}$. The emission is brighter on the northern side of the radio shell and decreases toward the south, where a break-out morphology is observed.
Based on data acquired with H.E.S.S., \citet{hof11} report the discovery of the TeV source HESS~J1818$-$154 (size 8$^{\prime}$.5) in spatial coincidence with the interior
of the SNR shell. The $\gamma$-ray source has a hard energy spectrum with a photon index
$\Gamma$=-2.1 and a total integral flux above 1~TeV of 
$\sim$4.0 $\times$ 10$^{-13}$~cm$^{-2}$~s$^{-1}$. 
On the basis of the morphological correspondence between the brightest hotspot of 
HESS~J1818$-$154 and the inner part of radio emission from G15.4+0.1, the authors propose that the high energy radiation originates in a pulsar wind nebula (PWN) of a still undetected pulsar.

To investigate the origin of the TeV $\gamma$-ray emission, we undertook a combined study that includes new observations of the radio continuum emission associated with G15.4+0.1, a search for  pulsations from a possible pulsar, and a characterization of the surrounding interstellar medium (ISM).
In this Letter, we present a new radio continuum image of G15.4+0.1 at 1.4~GHz obtained using the Giant Metrewave Radio Telescope (GMRT, India) and the first study of the atomic and molecular gas distribution in the vicinity of G15.4+0.1. The search for pulsations will be reported elsewhere.
We also determine the distance to the remnant and investigate a possible hadronic scenario to explain the high energy emission in the G15.4+0.1/HESS~J1818$-$154 system.

\section{Data}
The continuum image of G15.4+0.1 was constructed using new data acquired with the GMRT on 2012 May 10 and 11. To recover the 
large-scale structure missing in the interferometric image, we added single-dish data from the  Effelsberg~100m telescope.
We employed a standard procedure to reduce the data using the NRAO Astronomical Image Processing System (AIPS) package.
The resulting image has a sensitivity of 0.11~mJy~beam$^{-1}$ with a synthesized beam 
of 3$^{\prime\prime}$.5$\times$3$^{\prime\prime}.0$. A detailed study of the radio emission from G15.4+0.1 will be reported
separately.

The molecular  $^{13}$CO (J=1$-$0) line data come from the Galactic Ring Survey (GRS) 
\citep{jac06} while the 21~cm HI observations were taken from the Southern Galactic Plane 
Survey (SGPS) \citep{mcc05}. On the basis of these data, we have identified 
molecular material that we propose is physically associated with SNR~G15.4+0.1.

\section{Results}
\begin{figure*}[t!]
\begin{center}   
\includegraphics[scale=1, bb=37 290 576 470]{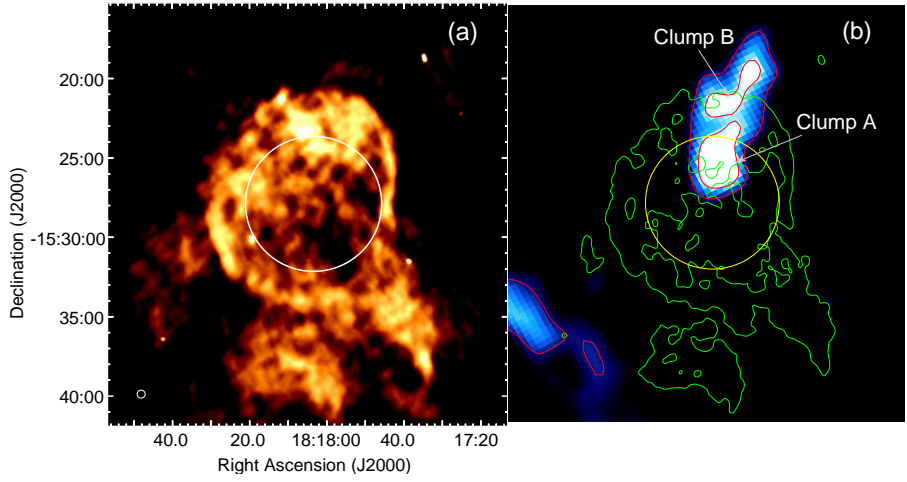}
\caption{\bf (a) \rm GMRT image of G15.4+0.1 at 1420~MHz smoothed to 15$^{\prime\prime}$ resolution, 
\bf (b) \rm Integrated channel map of $^{13}$CO (J=1$-$0) from GRS 
data over LSR velocities 46-50.3~km~s$^{-1}$. 
The circle marks the size and 
position of the $\gamma$-ray source HESS~J1818$-$154. The 0.45~mJy~beam$^{-1}$
contour of the 1420~MHz image is plotted for reference.}
             \label{co}
\end{center} 
   \end{figure*}

In Fig.~\ref{co} we present the distribution of the molecular gas integrated from 46.0 to 50.3~km~s$^{-1}$, the range 
where the CO emission is more intense (LSR radial velocities), 
with superposed contours of the radio emission at 1420~MHz. The left hand frame displays the 1420~MHz continuum
image of G15.4+0.1 for comparison.  

The analysis of the whole $^{13}$CO (J=1$-$0) data cube shows evidence of a nonuniform density distribution of the ISM in the region around G15.4+0.1. We identify two clumps of molecular gas, named in what follows as clumps A and B, which are 
apparently part 
of the same cloud (see Fig.~\ref{co}b). The molecular clump A, centered at 
R.A.(J2000)$\simeq 18^{\mathrm{h}}\,18^{\mathrm{m}}\,02^{\mathrm{s}}$, 
dec.(J2000)$\simeq -15^{\circ}\,25^{\prime}\,23^{\prime\prime}$, appears projected onto
the TeV source, while clump B at the position
R.A.$\simeq 18^{\mathrm{h}}\,17^{\mathrm{m}}\,57^{\mathrm{s}}$, 
dec.$\simeq -15^{\circ}\,20^{\prime}\,13^{\prime\prime}$
coincides with the bright northern portion of the remnant. In what follows we analyze whether
the high energy emission is related to these two clumps.

\begin{figure}[h]
\includegraphics[scale=0.7]{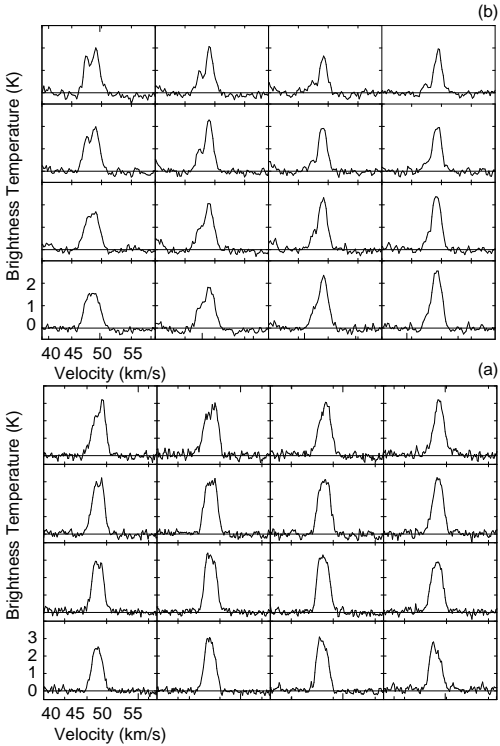}
\caption{$^{13}$CO (J=1$-$0) spectra observed around the center of the \bf (a) \rm clump A  and
\bf (b) \rm clump B.}
             \label{spectra-co} 
    \end{figure}

\begin{figure}[h]
\includegraphics[scale=0.7]{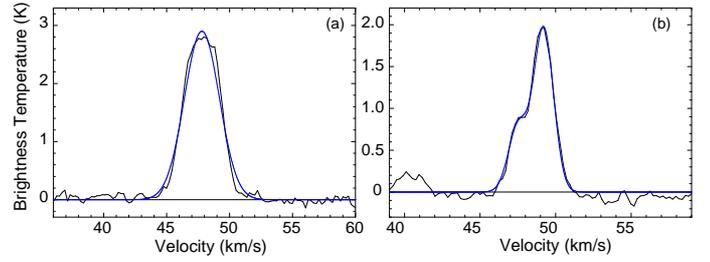}
   \caption{Averaged $^{13}$CO (J=1$-$0) spectra obtained toward \bf (a) \rm clump A and \bf (b) \rm clump B, in
the northernmost and central regions of the SNR~G15.4+0.1. The Gaussian fit to each spectrum
is shown by the blue curves.}
             \label{co-averaged} 
    \end{figure}

The $^{13}$CO spectra observed toward clumps A  and B (in an area of $\sim$ 1.46 square arcminutes) are shown in Fig.~\ref{spectra-co}. 
The average spectrum for clumps A and B are displayed in Fig.~\ref{co-averaged} in which the Gaussian fit to each spectrum is indicated by the blue curve.
For clump A, the best  fit, made with a single function, yields a central velocity
v$_{\mathrm{c}}$ and line FWHM $\Delta$v of 47.80$\pm$0.80 and 6.52$\pm$0.75~km~s$^{-1}$, respectively.
For clump B the CO profile reveals a shoulder at ``blueshift'' velocities. 
This spectrum can be fitted by two Gaussian curves, with central velocities v$_{\mathrm{c}}$ and
line FWHM $\Delta$v of 49.21$\pm$0.65 and 6.12$\pm$0.84,
and 47.53$\pm$0.46 and 6.16$\pm$0.75~km~s$^{-1}$, 
respectively. 
The presence of two velocity components might be evidence of a molecular 
cloud disrupted by the passage of a SNR shock front. 
Similar kinematical features have been observed in several cases of SNR/molecular cloud interactions,
as for example in the remnants G349.7+0.2 \citep{dub04}, Kes~69 \citep{zho09}, and G20.0$-$0.2 \citep{pet13}.
Of course, an alternative explanation based on the presence of multiple molecular components along the line of sight cannot be entirely ruled out.

\subsection{HI, $^{13}$CO spectra and distance constraints}
\begin{figure}[h!]
\includegraphics[scale=0.5]{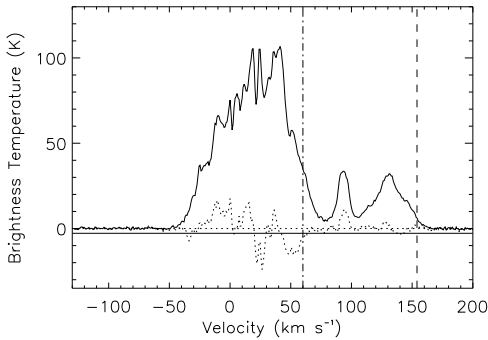}
 \caption{HI emission spectrum (solid curve), together with the absorption profile (dotted curve)
for the SNR~G15.4+0.1. The tangent point velocity is indicated by the dashed vertical line, while
the maximum velocity absorption feature is marked by the dot-dashed vertical line.}
             \label{hi-snr} 
    \end{figure}

To date there has not been a reliable estimate for the distance to SNR~G15.4+0.1.
From the Galactic $\Sigma$-$\Delta$ relation (surface brightness vs. diameter), \citet{hof11}
place the remnant at a distance of 10 $\pm$ 3~kpc.
However, the application of the $\Sigma$-$\Delta$ method to establish distance to individual SNRs
has been long questioned because of the severe uncertainties caused by its dependence on considerations 
about intrinsic and extrinsic factors related to the supernova explosion and the evolution of the remnant.
Here, in order to better constrain the distance to G15.4+0.1
we used HI absorption features seen against this SNR. 
In Fig.~\ref{hi-snr} we present the HI emission spectrum in the line of sight 
toward the brightest portion of the SNR shell, along 
with an absorption spectrum obtained
by subtracting the emission profile from a spectrum averaged over different regions 
adjacent to the remnant. The maximum velocity at which we observe absorption 
is $\sim$60~km~s$^{-1}$. According to the
circular rotation curve of the Milky Way by \citet{fic89} 
(assuming $\Theta$=220~km~s$^{-1}$, R$_{0}$=8.5~kpc),
this feature is associated with the near and far kinematical distances of $\sim$4.8 and $\sim$11.7~kpc, respectively.
The distance to the SNR can be further constrained by the fact that 
 only HI emission features are observed in the spectrum beyond $\sim$60~km~s$^{-1}$. If G15.4+0.1 were located farther away than 4.8~kpc, then one should observe absorption features at velocities greater than $\sim$60~km~s$^{-1}$ 
in the spectrum of the source  (because in the first quadrant of the Galaxy the velocity increases with distance up to the tangent point). As a consequence, we
conclude that G15.4+0.1 is located at the near distance of 4.8$\pm$1.0~kpc. 

To establish a physical association between the SNR and the $^{13}$CO molecular material we also determined the 
kinematical distance to the cloud containing clumps A and B 
to be $\sim$4.2 and $\sim$12.3~kpc, for the near and far distances.
To unambiguously establish the distance to the cloud,  
we used the $^{13}$CO (J=$1-0$) observations in conjunction with HI data. 
The technique takes  the atomic HI hydrogen within the cold interior 
of molecular clouds into account \citep{rom09}. As shown in Fig.~\ref{co}, both clumps are embedded in the radio continuum emission from the remnant. Usually, the brightness temperature of the 21~cm continuum radiation is higher than the temperature of the cold HI coexisting with molecular clouds.
If the clumps A and B are located at the near kinematical distance, all
the foreground molecular clouds along the same line of sight absorb the continuum radiation. 
Consequently, features will be observed in the HI 21~cm spectrum coincident with $^{13}$CO emission lines up to the
velocities of the clumps. 

\begin{figure*}[t]
   \centering
\includegraphics[scale=0.8]{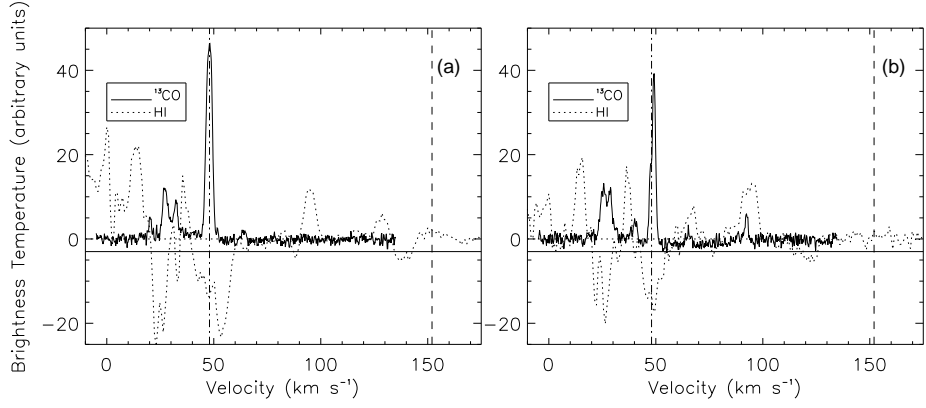}
  \caption{The $^{13}$CO and HI 21~cm spectra toward  \bf (a) \rm clump A and \bf (b) \rm clump B
embedded in the continuum emission from the SNR~G15.4+0.1. The vertical dash-doted line indicates
the central velocities of each clump, while the tangent point velocity is marked by a vertical dashed line. The solid horizontal line represents the error of about 2.8~K in the HI 21~cm absorption spectrum.}
             \label{hi-co} 
    \end{figure*}

Conversely, if clumps A and B lie at the far kinematical distance,
the foreground molecular clouds absorb the 21~cm continuum radiation embedded within the clumps at velocities up to the velocity of the tangent point, the maximum velocity that can have the foreground clouds. 
Therefore, the HI 21~cm spectrum toward the clumps will exhibit absorption lines 
corresponding to a $^{13}$CO emission line from the foreground molecular clouds
up to the velocity of the tangent point.
Figure~\ref{hi-co} displays the HI 21~cm and $^{13}$CO spectra in the 
direction of clumps A and B. 
For both clumps, the HI 21~cm spectrum shows absorption lines that correspond to molecular emission lines up to the central velocities derived for each clump. 
Therefore, we can assign the near kinematical distance (4.2$\pm$1.2~kpc) to both clumps. 
The weak absorption features at $\sim$140~km~s$^{-1}$ (clump A) and at $\sim$125~km~s$^{-1}$
(clump B) are of very low significance since they have the same magnitude as does  the uncertainty that we 
adopted in the HI absorption spectra ($\sim$2.8~K). 
We thus provided  a reliable estimate for the first time for the distance to the SNR and proved that 
the molecular clouds are located at the same distance.

\subsection{The SNR~G15.4+0.1/HESS~1818$-$154 system}
G15.4+0.1 is among the faintest SNRs  detected yet at TeV energies. Up to the present, 
it has been suggested that the discovered $\gamma$-ray emission originates in a PWN of a still undetected 
rotating neutron star \citep{hof11}.
In this section we attempt to use the $^{13}$CO  data to infer the physical properties 
of the molecular material that we suggest is interacting with SNR~G15.4+0.1. 
We adopt a distance of 4.2~kpc to estimate the total mass for the molecular clump
and its density. All the subsequent quantities are scaled in terms of d$_{4.2}$=d/4.2~kpc.

For clump A we consider the $^{13}$CO intensity from a circular region of about
0$^{\circ}$.06 in diameter.
We use the  $^{13}$CO distribution to calculate the H$_{2}$ 
column density by adopting the molecular mass calibration ratio 
N(H$_{2}$)/N($^{13}$CO)=5.62 $\times$ 10$^{5}$ \citep{sim01}, where 
N($^{13}$CO) is obtained from the following relation  \citep{wil09}

\begin{equation}
\mathrm{N}(^{13}\mathrm{CO})=2.42 \times 10^{14}\; \frac{(T_{\mathrm{ex}}+0.88)\,
\int{\tau_{13}\,dv}}{1-e^{-5.29/T_{\mathrm{ex}}}},
\label{n13}
\end{equation} 

\noindent
where T$_ {\mathrm{ex}}$ represents the excitation temperature of the 
$^{13}$CO (J=1$-$0) transition, which is assumed to be $\sim$ 10~K, and $\tau$$_{13}$ is the 
optical depth of the line. For an optically thin line, the integral in Eq.~\ref{n13} can 
be expressed as $\int{\tau_{13}\; dv} \sim \, \frac{1}{J(T_{\mathrm{ex}}) - J(T_{\mathrm{b}})} \,\int{T_{\mathrm{B}}} \; dv$, 
with $J(T)$=$5.29 \, (e^{5.29/T}-1)^{-1}$, T$_{\mathrm{B}}$ the brightness temperature of the line, and T$_ {\mathrm{b}}$ fixed at the typical value of 2.7~K. 

By using the estimation made for N(H$_{2}$) we obtain the mass for the central cloud through the relation,

\begin{equation}
M=\mu \, m_{\mathrm{H}} \, \Sigma \, [d^{2} \, \Omega \, N(\mathrm{H}_{2})],
\label{mass}
\end{equation}

\noindent
where  $\mu$ is the mean molecular weight equal to 2.8 if a relative helium abundance of 25\% is assumed, 
m$_{\mathrm{H}}$ represents the hydrogen mass, $d$ is the distance to the cloud, and $\Omega$ is the solid angle subtended by the $^{13}$CO cloud. 
We thus derive a total mass for the molecular clump A
of (1.2$\pm$0.5~$)\times$10$^{3}$~d$_{4.2}$$^{2}$~M$_{\odot}$.
The corresponding density  (for neutral hydrogen) 
is (1.5$\pm$0.4~$)\times$10$^{3}$~d$_{4.2}$$^{-1}$~cm$^{-3}$.

The volume density of the molecular gas can also be inferred by using an alternative method based on estimating the dust content.
Particularly, this clump matches the Bolocam\footnote{The Version 2.0 of the 
Bolocam GPS catalog is available at http://milkyway.colorado.edu/bgps/}
millimeter continuum source 
\object{BGPS~G015.433+00} about 0.035$^{\circ}$ in diameter. 
Following  \citet{ros10}, we analyze the 
properties of this source and estimate its mass from the flux density of the continuum emission
 at $\lambda$1.1~mm ($\nu$=271.1~GHz, $S_{\mathrm{\nu}}$=2.514$\pm$0.315~Jy). 
We obtain a  density  of (0.99$\pm$0.12)~$\times$10$^{3}$~d$_{4.2}$$^{-1}$~cm$^{-3}$ for clump A,
which agrees well with our estimate from  $^{13}$CO data. Similarly, the total mass of the northern clump B is estimated in 
 (3.0$\pm$0.7)$\times$10$^{3}$~d$_{4.2}$$^{2}$~M$_{\odot}$ from an elliptical area of axis 0$^{\circ}$.03 and 0.$^{\circ}$05, and the density is 
(1.1 $\pm$ 0.3)~$\times$10$^{3}$~d$_{4.2}$$^{-1}$~cm$^{-3}$. 

In what follows we analyze whether the $\gamma$-ray emission is related to the interaction
of the SN shocks with the molecular gas, particularly with clump A based on the good positional
match. It is possible that the emission of HESS~J1818$-$154 is produced through the decay of 
$\pi$$^{0}$ mesons created in the interaction between accelerated protons/nuclei and 
dense ambient gas. The required energy injected into accelerated hadrons 
to obtain the observed $\gamma$-ray 
flux can be estimated from the equation 
$W_{p}$ $\simeq$ $t_{pp\rightarrow \pi^{0}} \; L_{\gamma}$ \citep{aha07}, where 
$t_{pp \rightarrow \pi^{0}}$ $\simeq$ 4.5 $\times$ 10$^{15}$ $(\eta/cm^{-3})^{-1}$~s = 4.5 $\times$ 10$^{12}$~s is the
characteristic cooling time of protons through the $\pi$$^{0}$ production channel that corresponds to a mean proton
density $\eta$ $\sim$ 1000~cm$^{-3}$, and 
$L_{\gamma}(1-10 \, \mathrm{TeV})$=$4\, \pi\, d^{2}\,w_{\gamma}(1-10 \, \mathrm{TeV})$
$\simeq$3.2$\times$10$^{33}$~erg~s$^{-1}$
is the luminosity of the HESS source, where $w_{\gamma}$ is the $\gamma$-ray energy flux
between 1 and 10~TeV, and $d\sim4.2$~kpc is the adopted distance to the source.
Assuming that the power law energy spectrum, with index $\Gamma$=-2.1, extends up to energies of GeV \citep{aha05},
we find that the total energy of cosmic rays required to generate the observed $\gamma$-ray flux is 
$W_{p} \sim 4 \times 10^{48}$~erg.
Thus for G15.4+0.1, the conversion of a few percent of the 
explosion energy 
to the acceleration of protons up to $\geq$100~TeV
would be enough to explain the observed $\gamma$-ray flux in HESS~J1818$-$154 by 
nucleonic interactions in a medium with the calculated density.

\section{Conclusions}
Based on absorption techniques
we have estimated the kinematical distance to G15.4+0.1 in 4.8~kpc. 
This is significantly closer than the $\sim$10~kpc proposed by \citet{hof11} on the basis of 
the   $\Sigma$-$\Delta$ relation.
Additionally, we explored the radio emission and the ISM toward the SNR~G15.4+0.1 and proved for the first time 
that there is a roughly elliptical molecular cloud, about 5$^{\prime}$ in size, which consists of two bright clumps: clump A 
matching the location of the $\gamma$-ray emission from HESS~J1818$-$154 and clump B in coincidence 
with the northern border of the radio shell of the remnant, with masses of 
$\sim$1.2~$\times$10$^{3}$~M$_{\odot}$ 
and $\sim$3~$\times$10$^{3}$~M$_{\odot}$, and densities of 1.5$\times$10$^{3}$ and
1.1$\times$10$^{3}$~cm$^{-3}$, respectively. 
We demonstrated that the kinematical distances derived for the SNR~G15.4+0.1 and to the molecular
complex agree within errors and that the molecular cloud  
shows evidence of disruption
by a strong shock. 

Further energetic arguments confirm that a hadronic scenario is the most
likely one to explain the observed $\gamma$-ray emission. The detection
of $\gamma$-radiation in the MeV-GeV energy range would be a signature
for pion-generated gammas, providing beyond doubt the hadronic origin, while
on the other hand, the discovery of a pulsar associated with G15.4+0.1, would
favor a possible leptonic contribution. New searches for a pulsating source
are being conducted with the GMRT.

\begin{acknowledgements}
 This publication makes use of molecular line data from the Boston University-FCRAO Galactic Ring Survey (GRS). 
This research was partially funded by Argentina Grants
awarded by ANPCYT: PICT 0902/07, 0795/08 and 0571/11 and CONICET PIP: 2166/08, 0736/12. G.~C. and G.~D. are 
Members of the Carrera of Investigador Cient\'ifico of CONICET, Argentina. L.~S. is a PhD Fellow of CONICET, Argentina.
We thank the staff of the GMRT who have made these observations possible. GMRT is run by the National Centre for
Radio Astrophysics of the Tata Institute of Fundamental Research, India.
\end{acknowledgements}

\bibliographystyle{aa}  
\bibliography{bib-g15.4}
\IfFileExists{\jobname.bbl}{}
{\typeout{}
\typeout{****************************************************}
\typeout{****************************************************}
\typeout{** Please run "bibtex \jobname" to optain}
\typeout{** the bibliography and then re-run LaTeX}
\typeout{** twice to fix the references!}
\typeout{****************************************************}
\typeout{****************************************************}
\typeout{}
}

\end{document}